\documentclass[epjST]{svjour}

\usepackage[dvips]{graphicx}
\usepackage{dcolumn}
\usepackage{bm}
\usepackage{dsfont}
\usepackage{color}

\usepackage{epstopdf}

\usepackage[]{cite}

\usepackage{url}
\usepackage[colorlinks=true ,citecolor=blue,urlcolor=blue]{hyperref}
\usepackage{amsmath,amssymb}

\def\bs#1{\boldsymbol{#1}}
\def\txt#1{\textnormal{#1}}

\begin{document}


\title{Identifying topological edge states in 2D optical lattices using light scattering}

\author{Nathan Goldman\inst{1}\fnmsep\thanks{\email{ngoldman@ulb.ac.be}} \and J\'er\^ome Beugnon\inst{2} \and Fabrice Gerbier\inst{2}}
\institute{Center for Nonlinear Phenomena and Complex Systems - Universit\'e Libre de Bruxelles (U.L.B.), B-1050 Brussels, Belgium \and Laboratoire Kastler Brossel, CNRS, ENS, UPMC, 24 rue Lhomond, 75005 Paris}

\abstract{
We recently proposed in a Letter [\emph{Physical Review Letters {\bf 108} 255303}] a novel scheme to detect topological edge states in an optical lattice, based on a generalization of Bragg spectroscopy. The scope of the present article is to provide a more detailed and pedagogical description of the system -- the Hofstadter optical lattice -- and probing method. We first show the existence of  topological edge states, in an ultra-cold gas trapped in a 2D optical lattice and subjected to a synthetic magnetic field.  The remarkable robustness of the edge states is verified for a variety of external confining potentials. Then, we describe a specific laser probe, made from two lasers in Laguerre-Gaussian modes, which captures unambiguous signatures of these edge states. In particular, the resulting Bragg spectra provide the dispersion relation of the edge states, establishing their chiral nature. In order to make the Bragg signal experimentally detectable, we introduce a ``shelving method", which simultaneously transfers angular momentum and changes the internal atomic state. This scheme allows to directly visualize the selected edge states on a dark background, offering an instructive view on topological insulating phases, not accessible in solid-state experiments. }

\maketitle

\section{Introduction}  

Ultra-cold atoms in highly controllable coupling fields constitute a novel experimental tool for studying the rich many-body physics arising in two dimensions ~\cite{Lewenstein:2007,bloch2008,cooper2008a,dalibard2011a}. Motivated by the possibility of reaching interesting quantum phases, synthetic magnetic fields \cite{lin2009b} and spin-orbit couplings \cite{lin2011a,Wang:2012,Cheuk:2012} have been realized experimentally for neutral atoms. Today, the engineering of these synthetic gauge potentials opens an important path for the exploration of topological phases, such as quantum Hall (QH) states, topological insulators and superconductors, in the clean and versatile environment offered by cold-atom setups \cite{bloch2008,cooper2008a}.

For the last decades, these topological phases have gained the interest of the scientific community for their unique properties, such as quantized conductivities, dissipationless transport and edge-states physics \cite{HasanKane2010,qi2011a}. These impressively robust phenomena rely on an important concept, the so-called \emph{bulk-edge correspondence} \cite{Hatsugai1993,Qi2006}. Topological phases of matter are characterized by robust, integer-valued topological invariants related to the \emph{bulk} structure of the material. The bulk-edge correspondence stipulates that well-defined \emph{edge} excitations localized near the boundaries of the system are associated to these topological invariants. Such edge excitations are of tremendous practical importance, as they usually carry some form of current protected against perturbations as long as the topological structure is preserved. As such, they are at the origin of the dissipationless transport observed for these phases. For example, in the QH effect taking place in 2D electronic systems, the topologically invariant Chern number \cite{thouless1982a,Kohmoto1989a} guarantees the presence of current-carrying edge states, and imposes their chirality \cite{Hatsugai1993}.

Cold-atom realizations of topological phases therefore constitute a complementary, but also intrinsically appealing, playground to further deepen our understanding of these topological properties. However, the detectability of topological phases remains a fundamental issue in the cold-atom framework \cite{Sorensen:2005,Goldman:2007,Hafezi:2007,Palmer:2008,umucalilar2008a,Goldman2010a,StanescuEA2010,Liu2010a,Rosenkranz:2010,Bermudez:2010prl,Bermudez:2010,Bercioux:2011,alba2011a,Zhao2011,Kraus2012,Price2012,Buchhold2012,Dellabetta2012,Goldman:2012njp,Dauphin:2012,Yao:2012}, where transport measurements constitute a possible, but very challenging task \cite{Brantut:2012}. In this sense, alternative signatures of topological phases, together with novel experimental probes, have to be considered in this new context. Following this strategy, several schemes have been described to directly measure topological invariants,  based on spin-resolved time-of-flight \cite{alba2011a,Goldman:2012njp} and density measurements \cite{umucalilar2008a,Zhao2011}. Alternatively, Bloch oscillations could also be performed to evaluate the Berry's curvature in 2D atomic systems \cite{Price2012}, which could then provide an estimation of the Chern number when integrated over the Brillouin zone. 

Inspired by the bulk-edge correspondence, it has also been suggested that topological edge states could be directly probed \cite{StanescuEA2010,Liu2010a,Goldman2012,Kraus2012,Goldman:2013}. For example, in the context of cold-atom QH insulators, a satisfactory signature of the non-trivial topological order would be obtained by probing the dispersion relation of QH edge states, thus demonstrating their chiral nature. \\

It is the aim of the present work to describe in detail such a realistic probe. We choose to analyze this detection scheme for an optical-lattice setup reproducing the Hofstadter model \cite{Hofstadter1976}, which is one of the simplest tight-binding lattice model exhibiting non-trivial Chern numbers \cite{thouless1982a,Kohmoto1989a} and topological edge states \cite{Hatsugai1993}. The experimental realization of this model, using cold atoms in optical lattices, is currently in development in several laboratories \cite{aidelsburger2011a,Jimenez2012a,struck2012a}, based on the proposals \cite{jaksch2003a,gerbier2010a}. We believe that our detection scheme could easily be  extended to any ultracold-atom setup emulating 2D topological phases. 
\\


\section{The Hofstadter optical lattice and topological edge states}

\begin{figure}[h!]
	\centering
	\includegraphics[width=1\columnwidth]{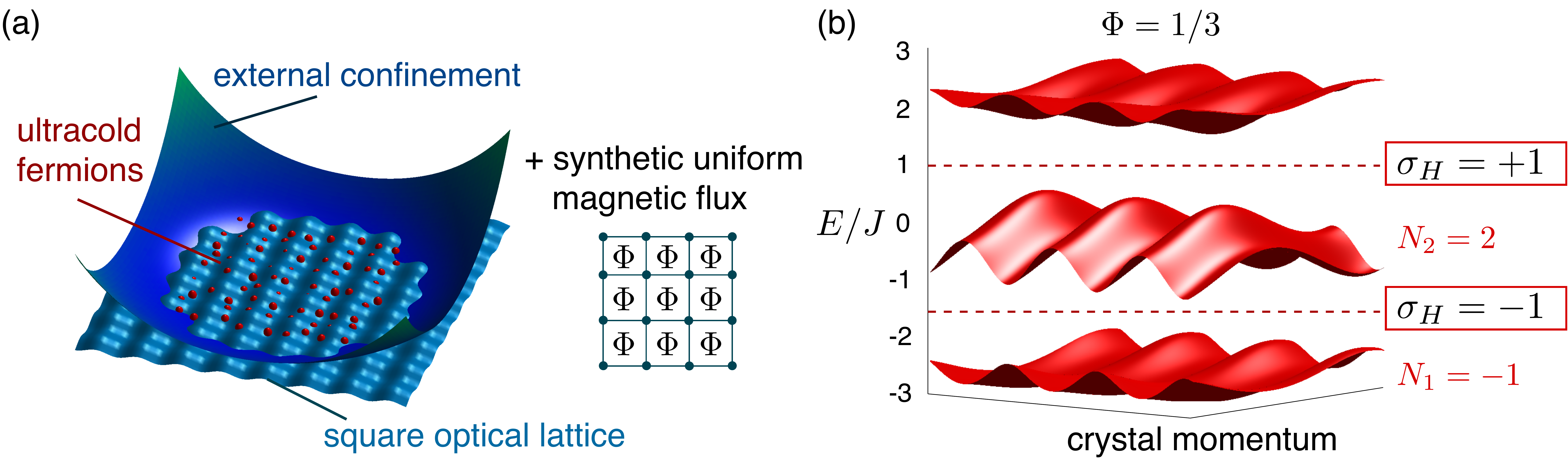}
	\caption{\label{figure1} (Color online) (a) The Hofstadter optical lattice: cold atomic fermions are trapped in a square optical lattice and subjected to a synthetic uniform magnetic flux $\Phi$. (b) The corresponding bulk energy spectrum $E=E(\bs k)$ for $\Phi=1/3$, and in the absence of external confining potential. The Chern numbers  $N_{1,2}$ associated with the two lowest bulk bands, and computed through Eq. \eqref{chern}, are shown. The Hall conductivity $\sigma_H$, computed from Eq. \eqref{Hall}, is indicated for Fermi energies $E_{\text{F}}$ located inside the two bulk gaps.}
\end{figure}

We start with a two-dimensional fermionic gas confined in a square optical lattice and subjected to a uniform synthetic magnetic field $\bs B=B \hat 1_z$ \cite{jaksch2003a,gerbier2010a}. The Hamiltonian is taken to be
\begin{align}
\hat H_0=& -J \sum_{m,n} \hat c^{\dagger}_{m+1,n} \hat c_{m,n} + e^{i 2 \pi \Phi m}  \hat c^{\dagger}_{m,n+1} \hat c_{m,n} + \text{h.c.} + \sum_{m,n} V_{\text{conf}}(r) \, \hat c^{\dagger}_{m,n} \hat c_{m,n},\label{ham}
\end{align}
where $\hat c_{m,n}$ is the annihilation operator defined at lattice site $(m,n) \in \mathbb{Z}^2$ and where $J$ is the tunneling amplitude. The site indices $m,n =1, \dots N$ are related to the spatial coordinates through $$(x,y)=a(m-N/2, n - N/2),$$ such that the center of the system is defined at $\bs x_{0} =(0,0)$. In the following, the lattice parameter $a$ defines our length unit. The external and circular confining potential is written as
\begin{equation}
V_{\text{conf}}(r) =J  \left(r/r_{\rm edge} \right)^\gamma , \label{confining}
\end{equation} 
where $\gamma=2$ ($\gamma=4$) corresponds to the standard harmonic (quartic) trap used in cold-atom experiments. The expression used for the confinement \eqref{confining}, including the tunneling amplitude $J$, is chosen such that the edge of the atomic cloud -- the Fermi radius -- is given by $R_{\text{F}} \approx r_{\rm edge}$, for the specific configuration considered in this work (i.e. $\Phi=1/3$ and $E_{\text{F}} \approx -1.5 J$, cf. Section \ref{sectionconf}). In the absence of the confinement $V_{\text{conf}}=0$, Eq.~(\ref{ham}) describes the Hofstadter lattice model, namely, a gas of non-interacting fermions in the tight-binding regime, subjected to a vector potential $\bs{A}=(0, B x,0)$ corresponding to $\Phi$ magnetic flux quanta per unit cell \cite{Hofstadter1976}. Methods to implement this optical-lattice setup, illustrated in Fig. \ref{figure1} (a), have been proposed in Refs. \cite{jaksch2003a,Sorensen:2005,gerbier2010a,Jimenez2012a}, and some important experimental steps towards this goal have already been successfully achieved \cite{aidelsburger2011a,Jimenez2012a,struck2012a}. \\ 

\begin{figure}[h!]
	\centering
	\includegraphics[width=1\columnwidth]{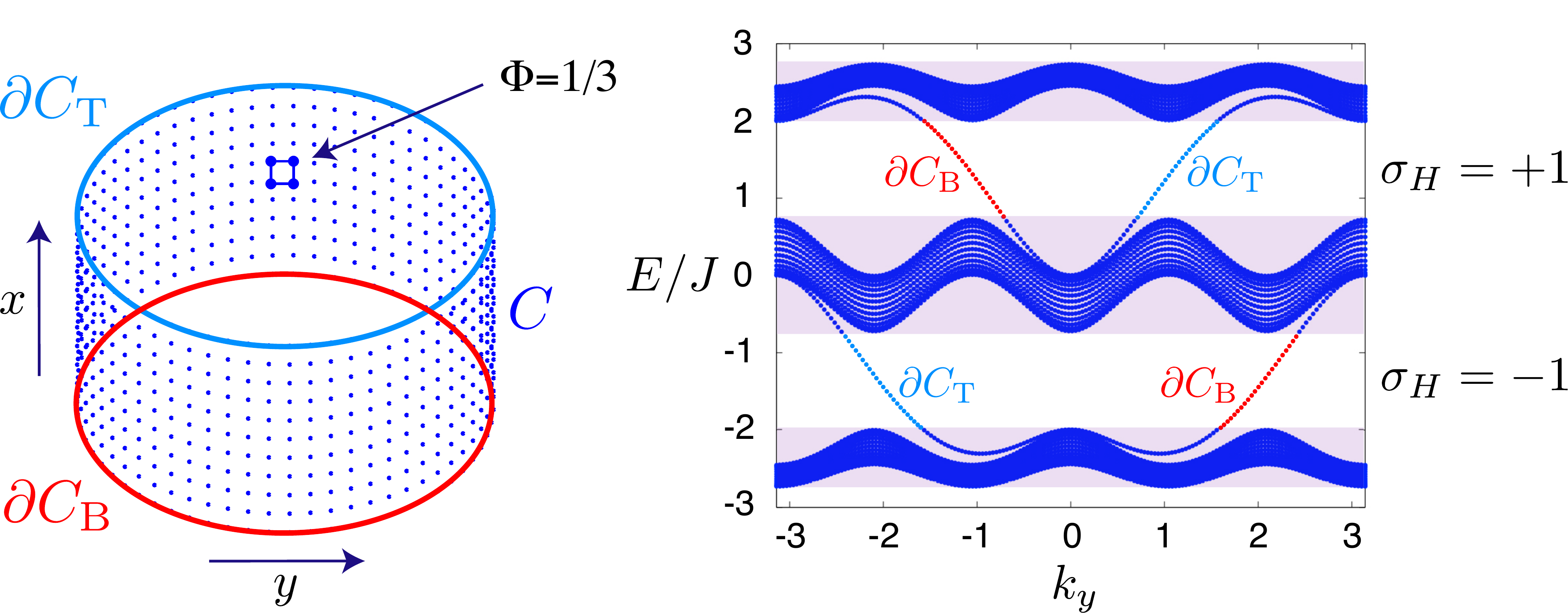}
	\caption{\label{figure2} (Color online) The cylinder analysis. (a) the Hofstadter optical lattice with $\Phi=1/3$ defined on an abstract cylinder $C$, in the absence of external confining potential. The two opposite edges of the cylinder are designated by $\partial C_{\text{T}}$ and $\partial C_{\text{B}}$, and an arbitrary unit plaquette is emphasized. (b) The corresponding energy spectrum $E=E(k_y)$, as a function of the quasi-momentum. The three bulk bands are designated by purple shades (see also the bulk spectrum in Fig. \ref{figure1}(b)). The dispersion branches located inside the bulk gaps correspond to chiral edge states, which are spatially localized at the two boundaries $\partial C_{\text{T,B}}$ of the cylinder. The edge states associated with different gaps have opposite velocities $v_e= E/\hbar k_y$.}
\end{figure}

\subsection{The topological edge states and the bulk-edge correspondence}

The transport properties and topological phases exhibited by the system can be deduced by solving the single-particle Schr\"odinger equation, which is performed through a direct diagonalization of the Hamiltonian \eqref{ham}. Setting $V_{\text{conf}}=0$ and considering periodic boundary conditions along the $x$ and $y$ directions -- namely, solving the system on a two-dimensional torus $\mathbb{T}^2$-- one obtains the energy band structure $E=E(\bs k)$ depicted in Fig. \ref{figure1} (b), where $\bs k=(k_x,k_y)$ is the quasi-momentum. For $\Phi = p/q$, where $p,q \in \mathbb{Z}$, the spectrum splits into $q$ subbands \cite{Hofstadter1976}, separated by bulk gaps. In the following, we set $\Phi=1/3$, in which case the spectrum depicts two large bulk gaps of the order $\Delta \sim J$. When the Fermi energy is set within a bulk gap, e.g. $E_{\text{F}} \approx \pm 1.5 J$, the interior of the system has the properties of an insulator. 

Because of the chosen periodic boundary conditions, the toroidal geometry cannot account for the edges present in real systems. In most physical systems, edge effects are normally neglected for sufficiently large sample size. However, topological phases constitute a counter-example, where the behavior at the edges represents an essential component of the physics. To see this, let us consider the same model on a cylindrical geometry, where periodic boundary conditions still hold in the $y$ direction, but where the system has a finite length in the $x$ direction. In this geometry, which is still abstract from the cold-atom point of view, the lattice system now features two edges, as represented in Fig. \ref{figure2} (a). The corresponding energy spectrum $E(k_y)$, shown in Fig. \ref{figure2} (b), can be partitioned in terms of \emph{bulk states} and \emph{topological edge states}. Indeed, one finds two dispersion branches within each bulk gap (cf. light blue and red curves in Fig. \ref{figure2} (b)), describing states that are spatially localized at the two edges of the cylinder \cite{Hatsugai1993}. Note that the dispersion relation is approximately linear in the middle of the gap, $E/\hbar \approx v_{e} k_y$ where $v_e$ is the group velocity, a feature generically found for such models \cite{qi2011a}. In particular, we find that the edge states located in different gaps have opposite chirality $\text{sign} (v_e)$. Therefore, when the Fermi energy is set within a given bulk gap, low-energy excitations propagate along the edge of the system with a specific chirality. In the solid-state framework, these chiral states are responsible for the quantum Hall effect \cite{HasanKane2010,qi2011a,Hatsugai1993}: when $E_{\text{F}} \approx \pm 1.5 J$, this propagating edge structure leads to the quantization of the transverse Hall conductivity \cite{Hatsugai1993}, $\sigma_H=\pm 1$ (in units of the conductivity quantum), as the two bulk gaps host a single edge-state branch per edge, with opposite chirality $\text{sign}(v_e)$. \\

In fact, these chiral edge states are \emph{topological}, in the sense that they are directly related to topological invariants associated with the bulk gaps. This important fact guarantees their robustness against small external perturbations: the edge states survive as long as the \emph{non-trivial} bulk gap in which they reside remains open. The fundamental concept which relates the edge states to topological invariants is the so-called \emph{bulk-edge correspondence} \cite{Hatsugai1993,Qi2006}, which is briefly described below.  First, let us introduce the Chern numbers $N_{\nu}$, which are topological indices defined for each bulk band $E_{\nu} (\bs k)$, labeled by the index $\nu$, through the Thouless-Kohmoto-Nightingale-Nijs expression (TKNN) \cite{thouless1982a},
\begin{equation}
N_{\nu}=  \frac{i}{2 \pi} \int_{\mathbb{T}^2} \langle \partial_{k_x} u_{\nu}(\bs k) \vert \partial_{k_y} u_{\nu}(\bs k) \rangle - (k_x \leftrightarrow k_y) \txt{d} \bs{k}.\label{chern}
\end{equation}
This formula corresponds to the integral over the first Brillouin zone ($\mathbb{T}^2$) of the Berry's curvature associated with the eigenstate $\vert u_{\nu}(\bs k) \rangle$ belonging to the band $E_{\nu} (\bs k)$. Here, $\bs k=(k_x,k_y) \in \mathbb{T}^2$ is the quasi-momentum. When the Fermi energy $E_{\text{F}}$ is exactly located in a bulk gap, the Hall conductivity is directly related to the Chern numbers,
\begin{equation} 
\sigma_H= \sum_{E_{\nu}<E_{\text{F}}} N_{\nu}, \label{Hall}
\end{equation} 
which can be derived from the Kubo formalism  \cite{thouless1982a,Kohmoto1989a}. Here the conductivity is expressed in units of the conductivity quantum. For the Hofstadter lattice  \eqref{ham} with $\Phi=1/3$, that we consider in this work, the bulk energy spectrum splits into three energy bands, which have the associated Chern numbers $N_1=-1$, $N_2= 2$ and $N_3 = -1$ (cf. Fig. \ref{figure1} (b)). Therefore, when the Fermi energy lies in the first (second) bulk gap, the Hall conductivity corresponds to $\sigma_H=-1$ ($\sigma_H=+1$), as illustrated in Fig. \ref{figure1} (b). In this sense, the Hall conductivity $\sigma_H=\pm 1$ is a non-trivial topological index characterizing the two bulk gaps \footnote{The Hall conductivity associated with a bulk gap can only change its value through a topological phase transition, that is, a gap-closing process.}. 

In such a non-trivial topological band configuration, the bulk-edge correspondence  dictates the following result \cite{Hatsugai1993,Qi2006}: if we solve the model \eqref{ham} on an open geometry, such as the cylinder considered above, gapless edge states will necessarily appear in the bulk gaps, because the latter are associated with non-zero topological invariants. Moreover the number of edge-state branches inside a bulk gap is given by the modulus of the Hall conductivity in \eqref{Hall}, and their chirality by $\text{sign}(\sigma_H)$. This fact is easily verified by comparing the numerical results shown in Figs. \ref{figure1} (b) and \ref{figure2} (b). The presence of a single edge excitation per physical edge in the lowest bulk gap of  Fig. \ref{figure2} (b) is in agreement with the fact that $\vert \sigma_H (\text{lowest gap}) \vert=\vert N_1\vert=1$: this is precisely the bulk-edge correspondence applied to the present context. 

\subsection{The Hofstadter optical lattice in an external confining potential}
\label{sectionconf}
\begin{figure}[h!]
	\centering
	\includegraphics[width=1\columnwidth]{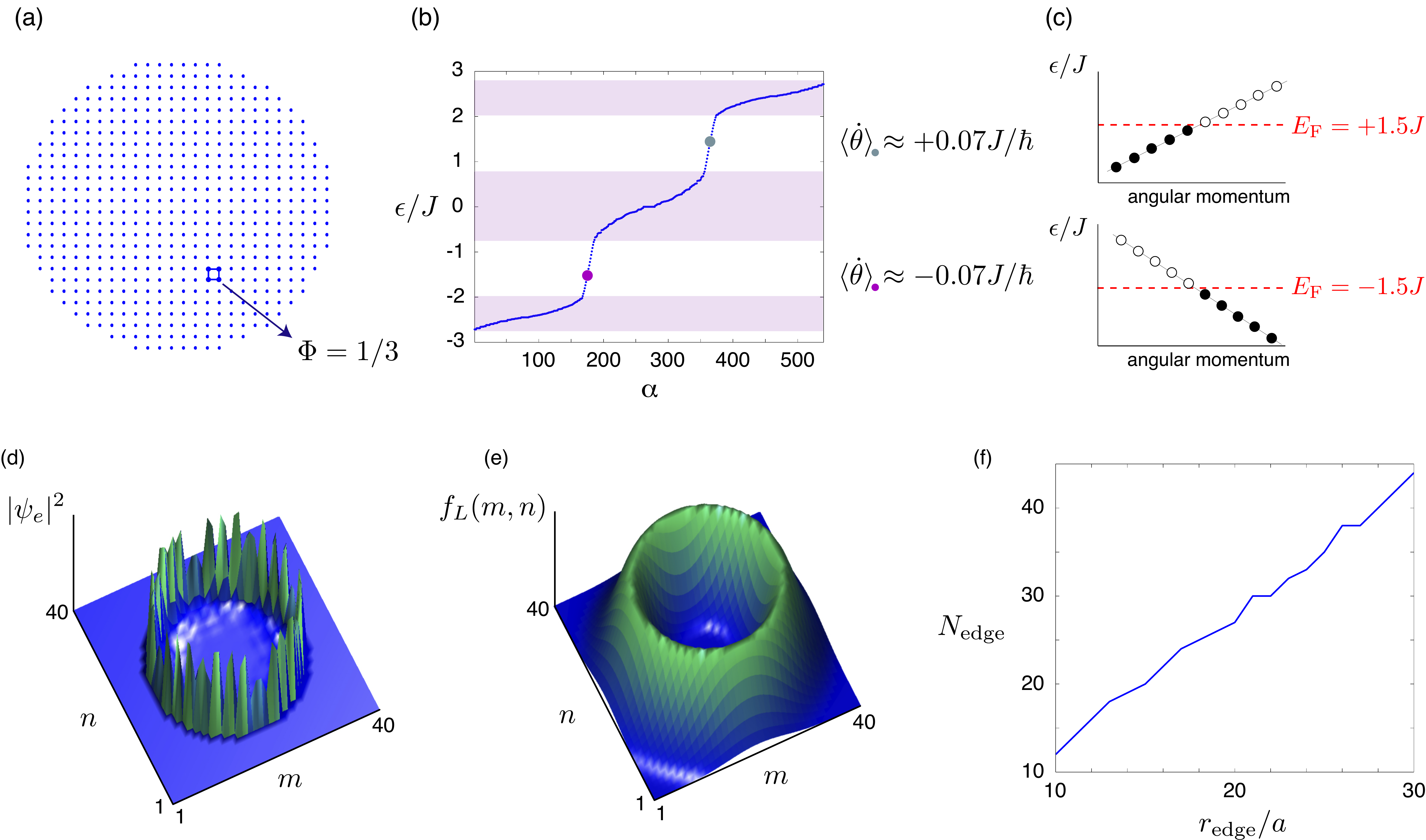}
	\caption{\label{figure3} (Color online) (a) The Hofstadter optical lattice in the cylindrically-symmetric confining potential $V_{\text{conf}} (r)$, with $\gamma=\infty$ and $r_{\text{edge}}\!=\!R_F\!=\!13 a$. A unit plaquette is highlighted on the figure. (b) The corresponding discrete energy spectrum $\epsilon_{\alpha}<\epsilon_{\alpha+1}$, where the index $\alpha$ classifies the eigenvalues in ascending order. The energy ranges corresponding to the bulk states are designated by purple shades (cf. Fig. \ref{figure1}). The angular velocity $\langle \dot{\theta} \rangle$ associated with the edge states at $\epsilon= \pm 1.5 J$, computed through Eq. \eqref{velo}, is shown in the gaps. (c) Sketch of the dispersion relations corresponding to the edge states around $\epsilon \approx \pm 1.5 J$. At zero temperature, all the states are occupied below the Fermi energy. (d) The amplitude $\vert \psi_e (m,n) \vert^2$ corresponding to the edge state $\vert \psi_{e} \rangle$ at $\epsilon_e \approx -1.5 J\!=\! E_{\text{F}}$, highlighted by a purple dot in (b). \!(e) The probe shape $f_L (m,n)\!=\!f_L (r)$ used to detect the edge state in (d).  (f) The number of edge states $N_{\text{edge}}$ inside the first bulk gap, as a function of the radius $r_{\text{edge}}$ for $\gamma = \infty$. For $r_{\text{edge}}=13 a$, shown in (b), one finds $N_{\text{edge}}=18$ edge states in the first bulk gap and $N_{\text{bulk}}=160$ states in the first bulk band: when $E_{\text{F}}=-1.5J$ and $r_{\text{edge}}=13 a$, there are $N_{\text{bulk}}=160$ occupied bulk states and only 8 occupied edge states (cf. discussion in Section \ref{shelvingsection}).}
\end{figure}

\begin{figure}[h!]
	\centering
	\includegraphics[width=1\columnwidth]{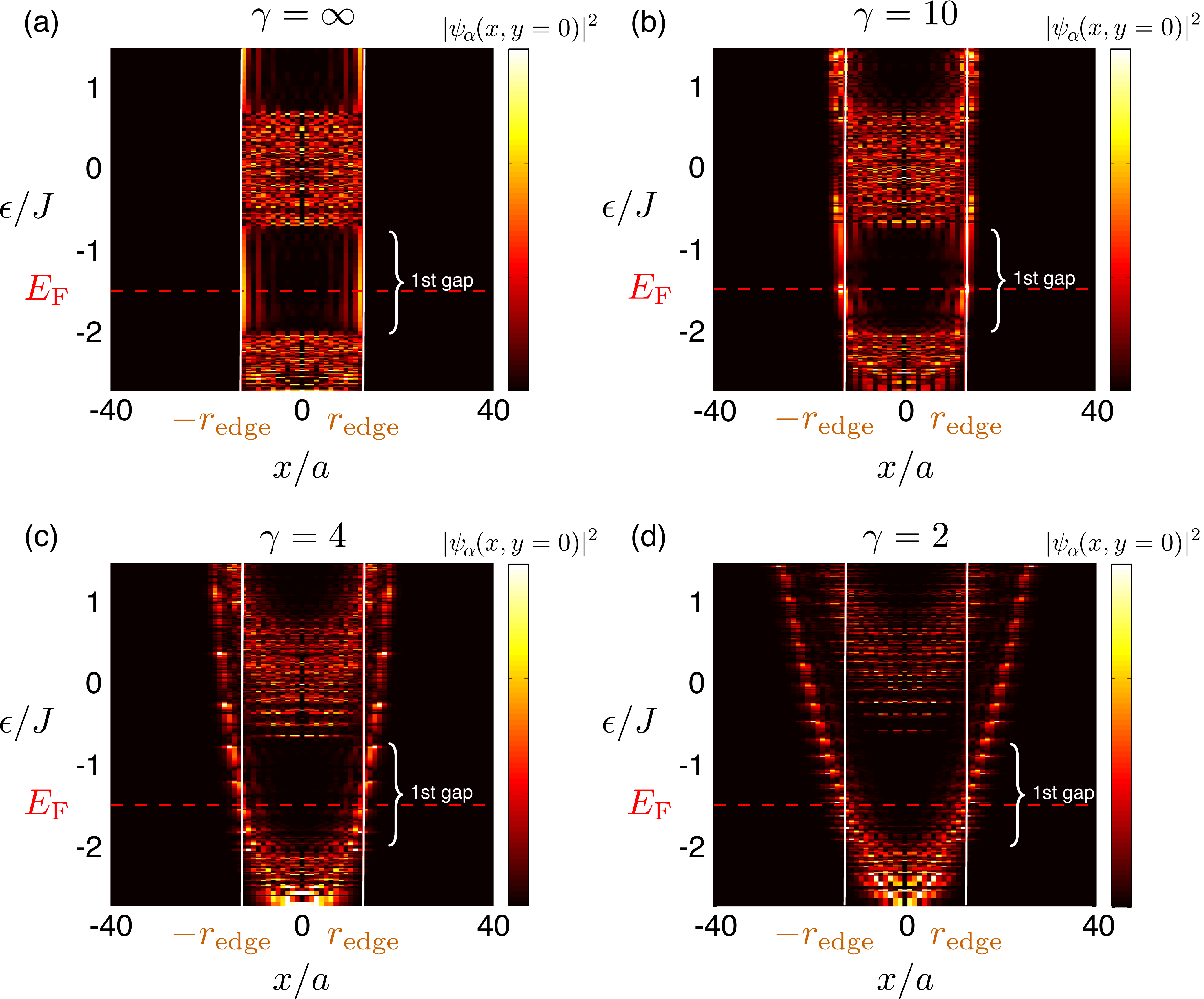}
	\caption{\label{figure4} (Color online) Confinement effect: The amplitudes $\vert \psi_{\alpha} (x,y=0) \vert^2$ associated with the single-particle states $\vert \psi_{\alpha} \rangle$ of the confined Hofstadter optical lattice (cf. Fig. \ref{figure3} (a)-(b)), as a function of their energy $\epsilon=\epsilon_{\alpha}$ and spatial coordinate $x$. Here the $y$ coordinate is chosen to be at the center of the trap, $y=0$. The confining potential is taken to be $V_{\text{conf}}(r)= J (r/r_{\text{edge}})^{\gamma}$, with $r_{\text{edge}}=13 a$, and (a) $\gamma = \infty$, (b) $\gamma = 10$, (c) $\gamma = 4$ and (d) $\gamma = 2$. The first bulk gap around $E_{\text{F}}=-1.5 J$, indicated on the figures, correspond to the energy range where chiral edge states are found. The radius $r=r_{\text{edge}}$, at which the Bragg probe is focused to scatter the edge states, is indicated by white vertical lines. Note that the Fermi radius $R_{\text{F}} \approx r_{\rm edge}$, for $E_{\text{F}} \approx -1.5 J$.}
\end{figure}

Having analyzed the edge-state structure, based on the cylinder analysis presented above, we now come back to the actual optical-lattice setup, whose finite size is determined by the confining potential $V_{\text{conf}}(r)$. For an infinitely abrupt potential ($\gamma=\infty$),
\begin{equation} 
V_{\text{conf}}( r \le r_{\text{edge}})=0 , \qquad V_{\text{conf}}(r > r_{\text{edge}})=\infty ,
\end{equation}
the lattice has the disk geometry represented in Fig. \ref{figure3} (a). The corresponding  spectrum and eigenstates are shown in Fig. \ref{figure3} (b) and Fig. \ref{figure4} (a). From the latter result, we find that the clear partition of the spectrum in terms of bulk states and topological edge states still holds in the experimental planar geometry: similarly to the cylindrical case, we obtain bulk states within the bulk bands and edge states (illustrated in Fig.~\ref{figure3}(d)) within the bulk gaps.  This result is in agreement with the bulk-edge correspondence, which requires that the chiral edge states lying inside the bulk gaps do not depend on the particular geometry of the lattice, as they are dictated by the Chern numbers \eqref{chern} associated with the bulk.  Therefore, when considering the realistic circular geometry produced by the confining potential $V_{\text{conf}}(r)$ in Eq. \eqref{confining}, one obtains the \emph{same} edge-state structure propagating along the circular edge $r=r_{\text{edge}}$ as the one obtained from the abstract cylinder discussed above:  the number of edge-state branches (per physical edge) and the chirality deduced from them are identical, as these properties do not depend on the chosen geometry.  In other words, the only difference between the cylindrical and the circular Hofstadter model is the number of physical edges (i.e. two edges for the abstract cylinder, and only one edge for the realistic circular geometry). However, let us comment on the fact that real boundaries, with finite $\gamma \ne \infty$, do affect the dispersion relations -- and thus the angular velocity -- of the edge states (cf. paragraph below). \\

The bulk-edge correspondence indicates that the lowest bulk gap in Fig. \ref{figure3} (b) hosts a single edge-state branch with a negative angular velocity, since the corresponding Hall conductivity $\sigma_H=- 1$ is solely governed by the topological expression \eqref{Hall}. Furthermore, the edge-state branch present in the second bulk gap corresponds to the opposite chirality, since $\sigma_H=+1$ when the Fermi energy is in the highest gap.  These results have been verified by directly computing the angular velocity of the edge states,  using the expression
\begin{align}
\dot{\theta}_{\text{e}}&=(i/\hbar) \langle \psi_e \vert [\hat{H}_0, \hat \theta] \vert \psi_e \rangle , \nonumber \\
&= \frac{i J}{\hbar} \sum_{m,n}   \bigl(  \theta(m+1,n) - \theta (m,n) \bigr ) \psi^*_{e} (m,n) \psi_{e} (m+1,n) \nonumber \\
&\qquad \qquad  + \bigl(  \theta(m-1,n) - \theta (m,n) \bigr) \psi^*_{e} (m,n) \psi_{e} (m-1,n)     \nonumber     \\
&\qquad \qquad+e^{i 2 \pi \Phi m} \bigl(  \theta(m,n+1) - \theta (m,n) \bigr ) \psi^*_{e} (m,n) \psi_{e} (m,n+1)  \nonumber \\
&\qquad \qquad+e^{-i 2 \pi \Phi m} \bigl(  \theta(m,n-1) - \theta (m,n) \bigr ) \psi^*_{e} (m,n) \psi_{e} (m,n-1),\label{velo}
\end{align}
where $\tan \theta (m,n)= (n-N/2)/(m-N/2)$, $m,n=1, \dots, N$, and where $\vert \psi_e  \rangle$ denotes a single-particle edge state close to the Fermi energy, $\epsilon_{e} \approx E_{\text{F}}$ (cf. Fig. \ref{figure3} (b) and (d)).  The numerical results $\dot{\theta}_{\text{e}} \approx \pm 0.07 J/\hbar$ (for $R_F \!=\!13a$) are shown in Fig. \ref{figure3} (b), together with a sketch of the corresponding dispersion relation in Fig. \ref{figure3} (c). \\

The partition of the energy spectrum in terms of bulk and edge states remains valid for finite confinements \cite{Buchhold2012}. This fact is demonstrated in Fig. \ref{figure4}(a)-(d) for different values of the parameter $\gamma=2,4,10$, defined in Eq. \eqref{confining}. In this figure, we observe how the edge-state structure smoothly follows the Fermi radius $R_{\text{F}}$, i.e., the edge of the atomic cloud imposed by the external confinement. In particular, for $E_{\text{F}} \approx -1.5 J$, we find that the edge states remain close to the parameter $r_{\text{edge}} \approx R_{\text{F}}$ defined in Eq. \eqref{confining}. In the following, we will consider a Fermi energy $E_{\text{F}}=-1.5~J$ located within the first bulk gap, so that the fermionic gas forms a QH insulator, with central density $n=1/3a^2$, and such that the edge states are located close to the radius $r_{\text{edge}}$. Let us finally emphasize that the edge states velocity highly depends on the boundary produced by the confinement: $\dot{\theta}_{\text{e}}$ significantly decreases as the potential $V_{\text{conf}}(r)$ is smoothened, \emph{e.g.}, $\dot{\theta}_{\text{e}} \approx-0.02 J/\hbar$ for $\gamma=10$ and $\dot{\theta}_{\text{e}} \sim -0.01 J/\hbar$ for $\gamma=2$ (for $R_F\!=\!13a$).

\section{Angular Momentum Spectroscopy :}

In the previous Section, we showed that the two bulk gaps at $\epsilon \approx \pm 1.5 J$ host topological edge states with \emph{opposite chirality}. The core of our proposal is to design an experimental probe, yielding a clear signature from these topological states, exploiting their specific chirality to distinguish them from the bulk. This probe is inspired by Bragg spectroscopy \cite{Liu2010a,StanescuEA2010}, a form of momentum-sensitive light scattering which is routinely performed to access the linear momentum distribution of cold atomic gases \cite{stenger1999b,steinhauer2002a}. First of all, we note that for the present problem, in which chiral edge states propagate along the circular edge of a 2D disk (cf. Fig. \ref{figure3} (a)), it is more convenient to probe the \emph{angular momentum} distribution in the vicinity of this edge\footnote{A standard Bragg spectroscopy, measuring the linear momentum distribution, could be used to probe the system as well. However, to achieve good overlaps with the edge states, and to avoid spurious signal from the bulk states, the Bragg lasers would have to be focused on a small region near the edge of the cloud, where one of the velocity's component, say $v_x$, remains approximately constant. Using simple dimensional arguments, one expects an extremely small excitation rate in this case. In contrast, our scheme exploiting angular momentum allows one to focus the probe lasers on the entire radius $r_{\text{edge}}$, maximizing the overlap with the edge states spatial distribution and thus the excitation rate.}. Therefore, we propose 

\begin{figure}[h!]
	\centering
	\includegraphics[width=0.4\columnwidth]{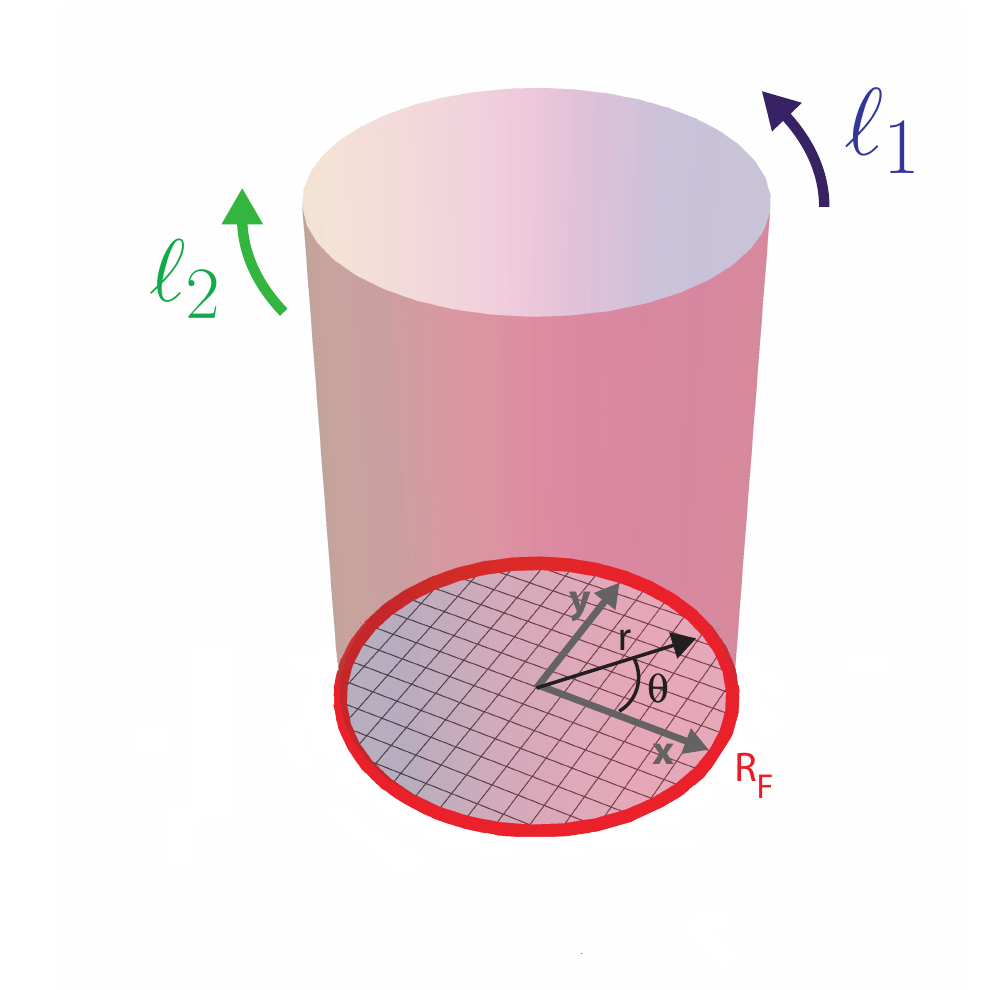}
	\caption{\label{figurejerome} (Color online) The Bragg probe configuration. The two Bragg beams, with opposite optical angular momenta $l_1$ and $l_2$, are represented by a shaded cylinder. These beams are incident on a 2D optical square lattice, where a cold Fermi gas is trapped within a Fermi radius $R_F$. Not shown is the coupling creating the synthetic magnetic flux.}
\end{figure}

\begin{itemize}
\item to use a spatial mode carrying angular momentum, in order to probe the angular momentum distribution;
 \item to shape the probing lasers to maximize (minimize) the probability to excite edge (bulk) states. 
\end{itemize}
We consider two lasers in high-order Laguerre-Gauss modes, denoted $1,2$, with optical angular momenta $l_{1,2}$, which correspond to the electric fields 
\begin{equation}
E_{1,2}(r) = \sqrt{I_{1,2}} f_{l_1,l_2} (r) \exp (-i l_{1,2} \theta - i \omega_{1,2} t), \quad f_{l}(r)\propto (r/r_0)^{\vert l \vert} e^{-r^2/2 r_0^2},
\end{equation} 
where $(r, \theta)$ are polar coordinates, cf. Fig. \ref{figurejerome}. The beams are assumed to be set off-resonance from a neighboring atomic transition, so that spontaneous emission can be neglected. This leads to a scattering Hamiltonian
\begin{align}
\hat H_{\text{Bragg}}(t) &=  \hbar \Omega \sum_{\alpha \beta} \bigl (  I_{\alpha \beta}^q e^{-i \omega_L t} +   I_{\alpha \beta}^{-q} e^{i \omega_L t} \bigr ) \hat c_{\alpha}^{\dagger} \hat c_{\beta},\\
I_{\alpha \beta}^q &=  \frac{1}{2} \int \text{d}\bs{x}\, \psi^{*}_{\alpha} (\bs{x})\psi_{\beta} (\bs{x}) f_L(r) e^{i q \theta}, \label{integrale}
\end{align}
where the index $q\!=\!l_2 - l_1$ represents the amount of angular momentum transferred by the probe (in units of $\hbar$),  $\hbar\omega_L\!=\!\hbar(\omega_1 - \omega_2)$ is the energy transfer,  and $\Omega$ is the Rabi frequency characterizing the strength of the atom-light coupling. The probe profile is   
\begin{equation}
f_{L}(r)\!=\!(r/r_0)^L e^{-r^2/r_0^2}/\mathcal{N}_L,
\end{equation} 
with $L\!=\!\vert l_1 \vert+\vert l_2 \vert$, and is illustrated in Fig. \ref{figure3} (e). The operator $\hat c_{\alpha}^{\dagger}$ creates a particle in the single-particle eigenstate $\vert \psi_{\alpha} \rangle$ of the unperturbed  Hamiltonian, i.e. $\hat{H}_0 \vert \psi_{\alpha} \rangle \!=\! \epsilon_{\alpha} \vert \psi_{\alpha} \rangle$.

Let us rewrite the scattering Hamiltonian as
\begin{align}
&\hat H_{\text{Bragg}}(t)= \hbar \Omega \bigl ( \hat{W}_q e^{-i \omega_L t} +  \hat{W}_{-q} e^{i \omega_L t} \bigr ) , \quad  \hat{W}_q=\sum_{\alpha \beta} I_{\alpha \beta}^q \hat c_{\alpha}^{\dagger} \hat c_{\beta}. \label{pert} 
\end{align}
Solving the time-dependent problem $\hat H_0+\hat H_{\text{Bragg}}(t)$ to first order, we write the many-body wave function as
\begin{align}
\vert \psi (t) \rangle &= b_0 (t) \vert 0 \rangle + \sum_{\mu} b_{\mu} (t) e^{-i E_{\mu} t/\hbar}  \vert \Psi_{\mu} \rangle \approx b_0 (t) \vert 0 \rangle + \sum_{(k,l)} b_{kl} (t) e^{-i \omega_{kl} t}  \vert k l \rangle,\label{wavedev}
\end{align}
where $\vert 0 \rangle=\prod_{\nu \le E_{\text{F}}} \hat c_{\nu}^{\dagger} \vert \emptyset \rangle$ denotes the groundstate at zero temperature, and 
\begin{equation}
\vert k l \rangle= \vert 1 \dots 1 \underbrace{0}_{l} 1 \dots 1 \underbrace{\vert}_{E_{\text{F}}} 0 \dots 0\underbrace{1}_{k} 0 \dots 0 \rangle, \label{myex}
\end{equation}
where $k > E_{\text{F}},l \le E_{\text{F}}$ and $\omega_{kl}=(\epsilon_k -\epsilon_l )/\hbar >0$. Here
we have restricted the full Hilbert space to the subspace spanned by the ground state and the excited states that are coupled to it to first order in the perturbation \eqref{pert}.

\begin{figure}[h!]
	\centering
	\includegraphics[width=1\columnwidth]{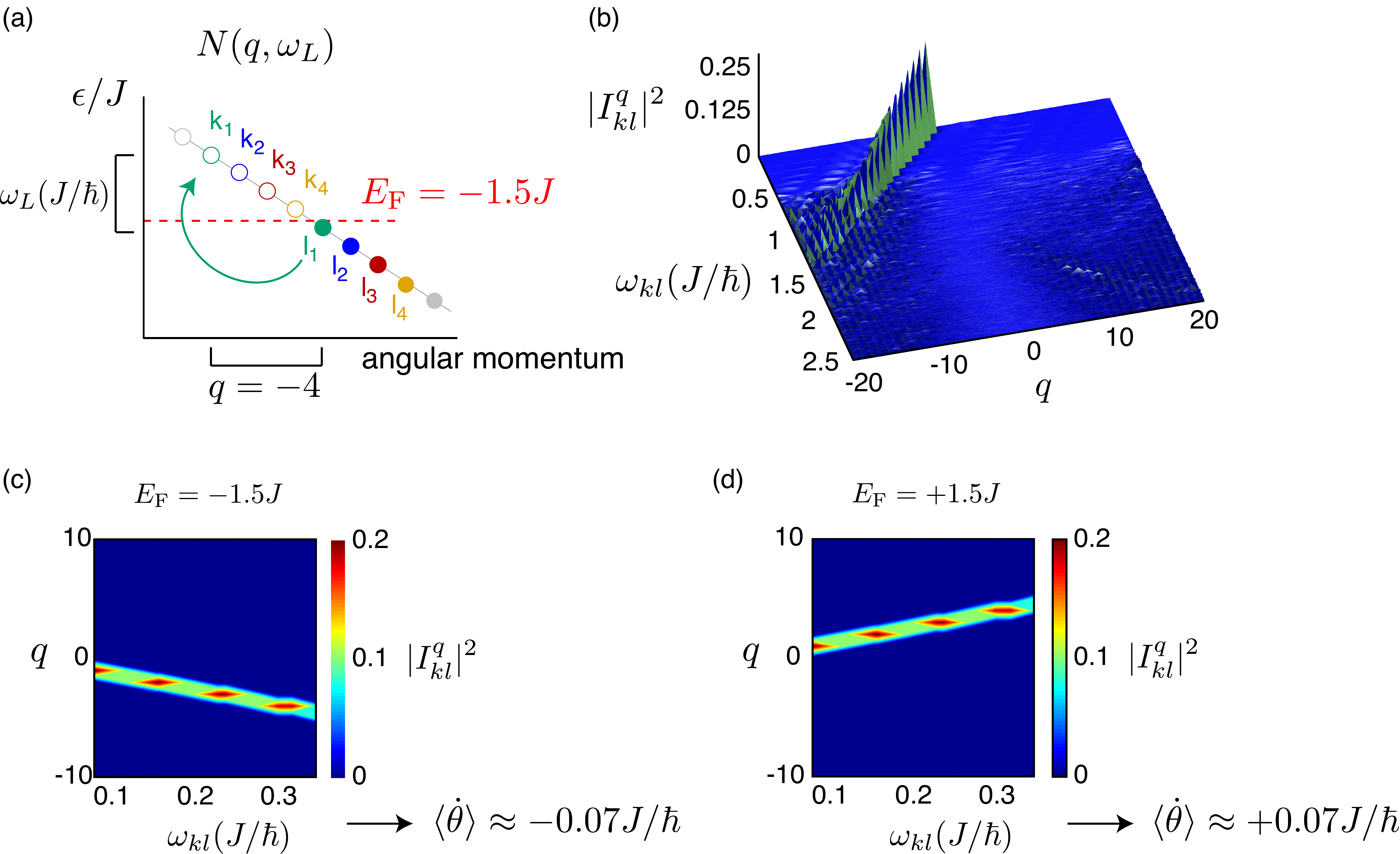}
	\caption{\label{figure5} (Color online) (a) Sketch of the single-particle energy spectrum $\epsilon_{\alpha}$ and the transitions $\vert \psi_{l} \rangle \rightarrow \vert \psi_{k} \rangle$ probed by $N(q, \omega_L)$, for $\omega_L \approx \omega_{kl} \ll J/\hbar$. (b) The amplitude $\vert I_{kl}^q\vert ^2$, as a function of the probe parameter $q$ and excitation frequency $\omega_{kl}$, for $\Phi=1/3$, $E_{\text{F}}=-1.5J$, $L=13$ and $r_0=5.1 a$. (c) The amplitude $\vert I_{kl}^q\vert ^2$, as a function of the probe parameter $q$ and excitation frequency $\omega_{kl}$, for  $E_{\text{F}}=-1.5J$ and (d) $E_{\text{F}}=+1.5J$. The confining potential is infinite ($\gamma = \infty$) and $r_{\text{edge}}=13 a$ in all figures. The angular velocity of the edge states present in the first [resp. second] bulk gap is $\dot{\theta}_{\text{e}} \approx \omega_{kl}^{\text{res}}/q \approx -0.07 J/\hbar$ [resp. $\dot{\theta}_{\text{e}} \approx +0.07 J/\hbar$]. Thus, the two bulk gaps are associated with opposite chiralities, in agreement with the result obtained from Eq. \eqref{velo}.} 
\end{figure}

Setting the initial condition $(b_0(0)=1, b_{kl}(0)=0)$, one finds 
\begin{equation}
b_{kl} (t)=- i \Omega  \biggl ( I_{kl}^q S_{kl}^- (\omega_L) e^{i \Delta_{kl}^{-} t }+(I_{lk}^q)^{*} S_{kl}^{+} (\omega_L) e^{i \Delta_{kl}^{+} t } \biggr ), 
\end{equation}
where $S_{kl}^{\pm} (\omega_L)= \sin (\Delta_{kl}^{\pm} t)/\Delta_{kl}^{\pm}$, $\Delta_{kl}^{\pm} =(\omega_{kl} \pm \omega_L)/2$. The number of scattered atoms, or excitation fraction, is then given by 
\begin{align}
&N(q,\omega_L) = \sum_{k,l} \vert b_{kl} (t) \vert^2 = \Omega^2 \sum_{k,l} \vert I_{kl}^q S_{kl}^- (\omega_L) e^{i \Delta_{kl}^{-} t }+(I_{lk}^q)^{*} S_{kl}^{+} (\omega_L) e^{i \Delta_{kl}^{+} t } \vert ^2. \label{finitetime}
\end{align}
In the long-time limit, and neglecting the anti-resonnant term $(\propto e^{i \Delta_{kl}^{+} t })$, this yields the standard Fermi golden rule
\begin{equation}
N(q,\omega_L)  = 2 \pi \Omega^2 t  \sum_{k > E_{\text{F}},l \le E_{\text{F}}} \vert I_{kl}^q \vert^2 \delta^{(t)}  (\omega_{kl} - \omega_L  ) , \label{fermiGR}
\end{equation}
where $\delta^{(t)}(\omega)\!= (2/\pi t) \sin ^2(\omega t/2)/\omega^2 \overset{t \rightarrow \infty}{\xrightarrow{\hspace*{1cm}}}   \delta(\omega)$. The expression \eqref{fermiGR} emphasizes the explicit relation between the excitation fraction $N(q,\omega_L)$ and the rates $\vert I_{kl}^q \vert^2$ defined in Eq. \eqref{integrale}. When the Fermi energy $E_{\text{F}}$ is set within a bulk gap, and for small frequency and  intensities $\omega_L,\Omega \ll J/\hbar $, the excitation fraction $N(q,\omega_L) $ probes the dispersion relation $\epsilon_e=\epsilon_{\text{e}} (M)$ associated with the gapless edge states $\vert \psi_e \rangle$ that lie within this gap, where $M$ is a quantum number analogous to angular momentum (see Fig.~\ref{figure5} (a)). For an optimized probe shape $f_L(r)$, which we obtain by setting $r_0 \approx r_{\text{edge}}/ \sqrt{\vert L \vert/2}$, this can be deduced from the behavior of the overlap integrals $I_{kl}^q$ defined in Eq.~\eqref{integrale}. They are represented in the $\omega_{kl}-q$ plane in Fig.~\ref{figure5}(b) for $\gamma= \infty$. At low frequencies $\omega_{kl} \ll J / \hbar$, we find a continuous alignment of resonance peaks $\omega_{kl}^{\text{res}}\approx \dot{\theta}_{\text{e}} q$. This reflects the linear dispersion relation $\epsilon/\hbar \approx \dot{\theta}_{\text{e}} M$ in the vicinity of the Fermi energy, and provides the angular velocity $\dot{\theta}_{\text{e}} \approx -0.07 J/\hbar$ and the chirality (i.e. $\text{sign}(\dot{\theta}_{\text{e}})$) characterizing the edge states in the lowest bulk gap. We find that this result is in perfect agreement with the direct evaluation of the angular velocity,  obtained through Eq.\eqref{velo}. As already mentioned, we stress that the edge states velocity  significantly decreases as the potential $V_{\text{conf}}(r)$ is smoothened. The absence of substantial response for $q>0$  in Fig.~\ref{figure2} (b) clearly proves that our setup is effectively sensitive to the edge state chirality. Naturally, the signal obtained by setting the Fermi energy in the second bulk gap, or by reversing the sign of the magnetic flux $\Phi \rightarrow - \Phi$, would probe the \emph{opposite} chirality. \\

\begin{figure}[h!]
	\centering
	\includegraphics[width=1\columnwidth]{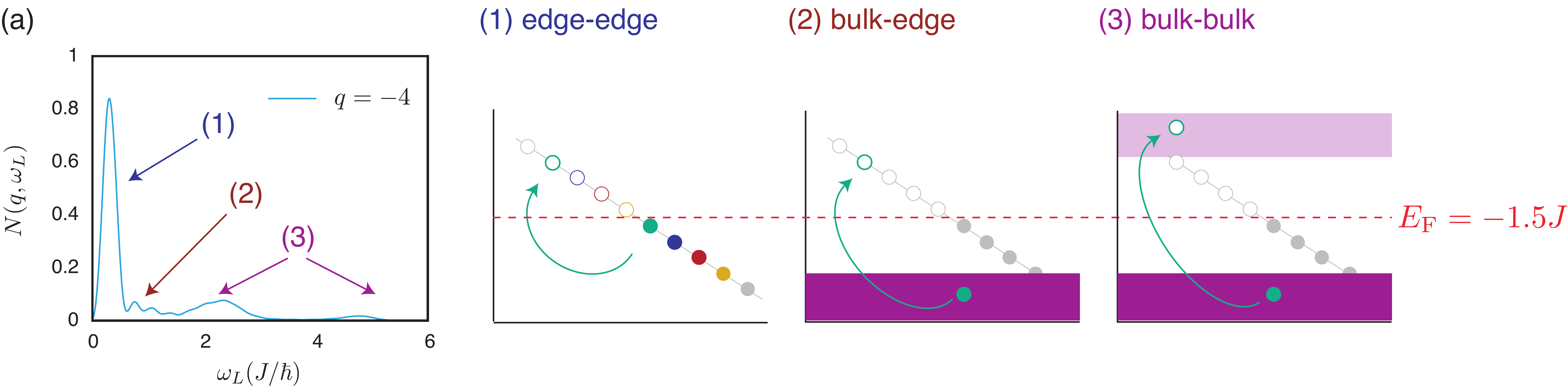}
	\caption{\label{figure6} (Color online) (a) Excitation fraction $N(q,\omega_L)$ versus probe frequency, for an angular momentum transfer $q\!= - 4$. Here, $\Omega\!=\!0.05 J/\hbar$, $t\!=\!20 \hbar/J$, $L\!=\!13$, $r_0\!=\!5.1 a$, $r_{\text{edge}}\!=\!13 a$, $\Phi\!=\!1/3$, $E_{\text{F}}\!=\!-1.5 J$ and $\gamma=\infty$.  The panels (1)-(3) sketch the three regimes of light scattering, shown in the main figure, for $q\!= - 4$: (1) Processes between edge states located in the gap around $E_{\text{F}}=-1.5J$. (2) Processes between the lowest bulk band and the edge states located above the Fermi energy. (3). Transitions between the two lowest bulk energy bands. }
\end{figure}

\begin{figure}[h!]
	\centering
	\includegraphics[width=1\columnwidth]{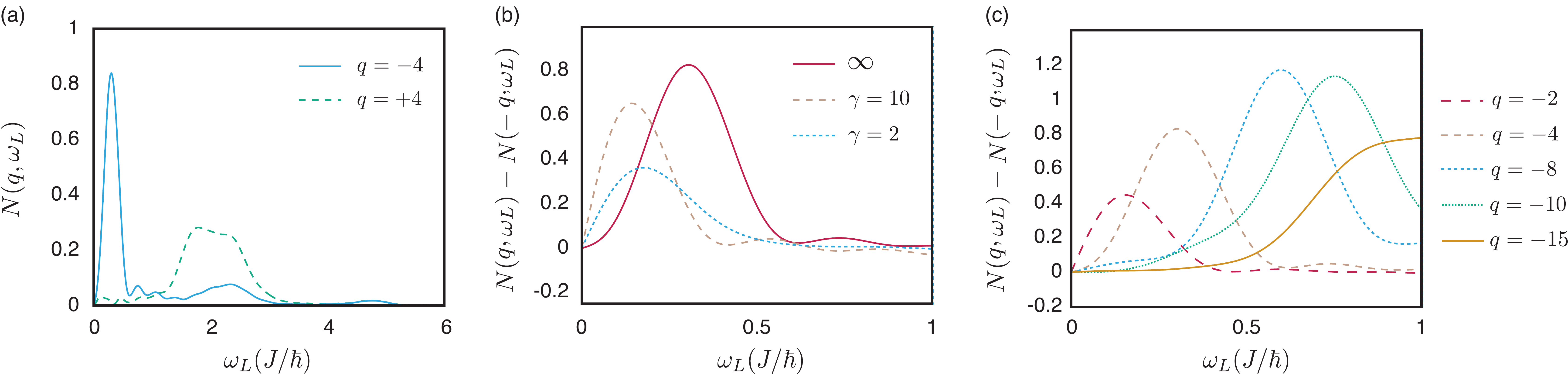}
	\caption{\label{figure7} (a) Excitation fraction $N(q,\omega_L)$ versus probe frequency, for an angular momentum transfer $q\!=\!\pm 4$. (b) $N(q,\omega_L)\!-\!N(-q,\omega_L)$ for several shapes of the trapping potential $V_{\text{conf}}(r) \!=\! J  \left(r/r_{\rm edge} \right)^\gamma $ and $q \!=\! -4$. (c) Increasing then broadening of $N(q,\omega_L)\!-\!N(-q,\omega_L)$, for increasing $\vert q \vert$. In all the figures $\Omega\!=\!0.05 J/\hbar$, $t\!=\!20 \hbar/J$, $L\!=\!13$, $r_0\!=\!5.1 a$, $r_{\text{edge}}\!=\!13 a$, $\Phi\!=\!1/3$, $E_{\text{F}}\!=\!-1.5 J$ and $\gamma=\infty$ (except in (b)).} 
\end{figure}

At finite times, it is preferable to evaluate the excitation fraction through a numerical evaluation of the Schr\"odinger equation,
\begin{align}
&i \hbar \frac{\text{d} b_{kl} (t)}{\txt{d} t}= \hbar \Omega \sum_{n,m} W_{kl;nm} (t) b_{nm} (t) e^{i (E_{kl}- E_{nm}) t /\hbar} , \label{numeric}
\end{align}
where $W_{kl;nm} (t)\!=\!\langle kl \vert \hat{W} (t) \vert  nm \rangle$ and $\hat{W} (t)= \hat{W}_q e^{-i \omega_L t} +  \hat{W}_{-q} e^{i \omega_L t}$. The many-body wavefunction is still restricted to the first-order subspace but off-resonant terms and deviations from the long-time limit are included. For the reasonable finite times and small Rabi frequencies $\Omega \ll J/\hbar $ used in our calculations, we find that the excitation fraction $N(q, \omega_L)$ obtained from a numerical resolution of Eq. \eqref{numeric} is in perfect agreement with Eq. \eqref{finitetime}.  A typical result is presented in Fig.~\ref{figure6}, for $q= - 4$, $t= 20 \hbar/J$ and $E_{\text{F}}=-1.5 J$, emphasizing the three distinct regimes of light scattering: ``edge-edge", ``bulk-edge" and ``bulk-bulk". The ``edge-edge" regime  corresponds to transitions solely performed between the edge states close to $E_{\text{F}}$: A sharp resonance peak is visible at $\omega_{L}^{\text{res}}\approx \dot{\theta}_{\text{e}} q\approx 0.3 J/\hbar$  for $q=-4$, and stems from four transitions between edge states, as sketched in Fig.~\ref{figure5}(a). Then, at higher frequencies, $\omega_L \approx J/\hbar$, small peaks witness allowed transitions between the lowest bulk band and the edge states located above $E_{\text{F}}$. Finally, for $\omega_L \approx 2 J/\hbar$, many transitions between the two neighboring bulk bands lead to a wide and flat signal. As shown in Fig. \ref{figure7} (a), this bulk-bulk response is significant for both $q=\pm 4$, as a consequence of the large density of excited states in this frequency range. In the following, we consider the quantity $N(q, \omega_L) -N(-q, \omega_L)$, which is zero for a system with time-reversal symmetry \cite{Goldman2012}. We have repeated the calculations for several potential shapes,  finding no qualitative change  (cf. Fig.~\ref{figure7}(b)). Although it is advantageous to use a steep confining potential, where the edge states are exactly localized at $r=r_{\text{edge}}$, the signal from the edge states is robust even in a harmonic trap ($\gamma=2$). We stress that a well focused probe allows to significantly reduce any signal from the bulk. Finally, we note that excitation times of several $\hbar/J$, which seem experimentally realistic, are long enough to resolve the edge-edge resonance but still too short to neglect the broadening due to the finite pulse time.  \\

As can be deduced from Fig.~\ref{figure5}(a), the number of allowed transitions $\vert \psi_{l} \rangle \rightarrow \vert \psi_{k} \rangle$ scales with the probe parameter $q$ in the ``edge-edge" regime. This is a consequence of the linear dispersion relation in the vicinity of $E_{\text{F}}$. Thus, one observes an increase of the peaks for increasing values of $\vert q \vert$ (cf. Fig. ~\ref{figure7}(c)). We stress that this progression only occurs in the ``edge-edge" regime, namely when $q$ is chosen such that $\hbar \omega_{L}^{\text{res}} \approx \hbar \dot{\theta}_{\text{e}} q$ is smaller than the energy difference between $E_{\text{F}}$ and the closest bulk band. In the case illustrated in  Fig. ~\ref{figure7}(c), the ``edge-edge" regime is delimited by $\vert q_{\text{e-e}} \vert \lesssim 7$. Beyond $\vert q_{\text{e-e}} \vert$, the resonance peak enters the ``bulk-edge" regime. In this case, the excitation fraction $N(q, \omega_L)$ broadens, $N(-q, \omega_L)$ is no longer negligible, and the linear dispersion relation is no longer probed. We thus conclude that a moderate value (here $\vert q \vert \sim4$) is preferable to keep a narrow peak, well separated from the broader ``edge-bulk" signal.

\section{Isolating and imaging the edge states: The Shelving method}
\label{shelvingsection}

\begin{figure}[h!]
	\centering
	\includegraphics[width=1\columnwidth]{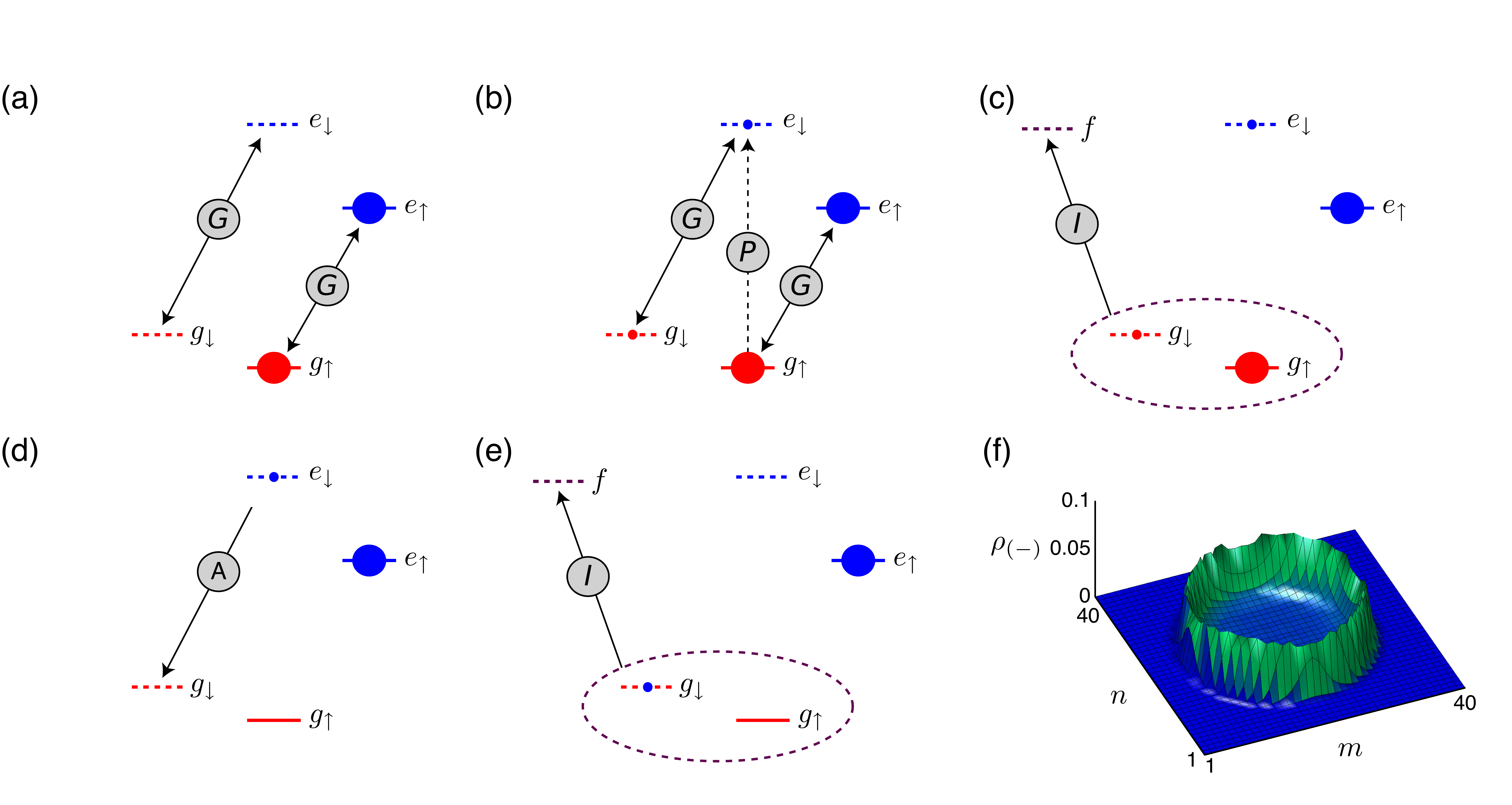}
	\caption{\label{figure8} (Color online) Shelving method to detect edge states. The scheme is presented for $^{171}$Yb,  a two-electron atom, with $g$ the electronic ground state manifold and $e$ a metastable excited state manifold, both with spin 1/2. (a)  The laser $G$ couples the states $g_\uparrow$ and $e_\uparrow$, but also $g_\downarrow$ and $e_\downarrow$, to generate the artificial magnetic flux in both $(+)$ and $(-)$ sectors (cf. main text). The states $g_\uparrow$ and $e_\uparrow$ are initially populated, while the other states are empty. (b) In order to probe the $(+)=\{g_\uparrow,e_\uparrow\}$ system, one introduces a weaker additional laser $P$, coupling $g_\uparrow \rightarrow e_\downarrow$. The circles indicate relative populations in each state after the probe pulse. (c) After the probe pulse, atoms in the $g$ manifold are dispatched using an auxiliary imaging transition $g \rightarrow f$,  where $f$ designates a second excited state with short radiative lifetime, suitable for applying a strong radiation pressure pulse and kick the $g$ atoms out of the trap. (d) The $e_\downarrow$ atoms are subsequently brought down to the $g$ manifold using, {\it e.g.}, adiabatic-passage techniques to ensure near-unit transfer efficiency. (e) The $e_\downarrow$ atoms are finally detected, without stray contributions from unperturbed atoms in other internal states. (f)   The local density $\rho_{-}$ that would be imaged through the Shelving method, cf. also Figs. \ref{figure9} (a)-(b) for a comparison between $\rho_{-}(+ q)$ and $\rho_{-}(-q)$.  Here, the excitation fraction is $N(q, \omega_L)=\sum_{m,n} \rho_{-}(m,n)$.}
\end{figure}

So far, we were able to show that a probe sensitive to orbital angular momentum, with a suitable spatial excitation profile, was able to detect the edge states and to establish their chiral character. With respect to experimental detection, the Bragg scheme presented above presents difficulties. The associated signatures in the spatial or momentum densities are small perturbations on top of the strong ``background" of unperturbed atoms, since the edge states represent a small fraction of the possibly available single-particle states. For a circular system with Fermi radius $R_F$, one can expect about $N_{\rm edge}\sim \Delta/\hbar \dot{\theta}_e$ edge states, where the energy gap $\Delta$ is fixed by the bulk problem. The angular velocity $ \dot{\theta}_e$ depends on the confining potential, as discussed above. We rewrite it as $ \dot{\theta}_e=\eta J a/ (\hbar R_F)$, with $\eta<1$ a numerical factor that depends on the trap potential and system size (e.g. $\eta\approx 1, ~0.3,~0.1$ for $\gamma=\infty,10,2$, respectively), $J a/\hbar$ a typical velocity associated with the band structure and $R_F$ the radius of the sample ($ R_F\sim a \sqrt{3N/\pi}$ for the regime considered in this paper: $E_{\text{F}}=-1.5 J$ and $\Phi=1/3$). Since the gap $\Delta \sim J$, one finds the geometrically intuitive relation  $N_{\rm edge}\sim R_F/a \eta$, indicating that the number of edge states scales as the ratio of the surface to the perimeter. Figure~\ref{figure3}(f) shows a numerical evaluation of $N_{\rm edge} (r_{\text{edge}})$, for the case $\gamma = \infty$ where $R_F=r_{\text{edge}}$.

The number of edge states present within the bulk gap sets an upper bound to the number of atoms that can be transferred by the probe.  Based on the argument above, and using the parameters from Fig.~\ref{figure3}(a), one finds $N_{\rm edge} \sim 13$ while the total atom number in the calculation is $N = 168$ (for $R_F = r_{\text{edge}}=13 a$ and $E_{\text{F}}=-1.5 J$). This estimation is in good agreement with the numerical result presented in Fig. \ref{figure3} (f).   Scaling to more realistic numbers for an experiment ($N\sim 10^4$, or $R_F/a\approx100$) leads to $N_{\rm edge}\sim 100/\eta$. This means that one should be able to detect a few tens of atoms at best on top of the signal coming from $\sim 10^4$ unperturbed ones: This is a significant experimental challenge with present-day technology. One possibility to avoid this difficulty is to use an alternate detection scheme, where the probe also changes the atomic internal state, thus allowing for independent detection of the coupled and perturbed atoms. The probe signal can then be measured against a dark background (without unperturbed atoms), which allows diverse and powerful imaging methods (such as large aperture microscopy, as recently demonstrated for quantum gases in optical lattices \cite{bakr2009a,sherson2009a}) to be used.

\subsection{Shelving detection method using state-changing transitions: example for $^{171}$Yb atoms}

We present in Fig.~\ref{figure8} a possible detection scheme suitable for two-electron atoms with ultra-narrow optical transitions, inspired by the electron shelving method  (in the following, we consider $^{171}$Yb atoms to give a specific example). The ground $g$ and metastable excited $e$ states have zero electronic angular momentum but nuclear spin $I=1/2$. We denote the Zeeman manifolds $\{g_\downarrow,g_\uparrow \}$ and $\{e_\downarrow,e_\uparrow \}$ in the ground $^1 S_0$ and  $^3 P_0$ excited states, respectively. The states $g_\uparrow$ and $e_\uparrow$ are initially populated, as laser coupling between these two states is used to generate the artificial gauge field \cite{jaksch2003a,gerbier2010a} leading to Eq.\eqref{ham}.  The strategy, which is at the core of our detection scheme, is based on the transfer of chiral edge states present in the populated sector $(+)=\{g_\uparrow,e_\uparrow\}$, to the empty sector $(-)=\{g_\downarrow,e_\downarrow\}$. A crucial point is to ensure that the topological edge states have the same structure in the initial and final states, so that a specific chirality is probed (cf. Section \ref{ana}).  To this end, the initially unpopulated  states $(-)$ are also coupled by a laser generating the same gauge field as for $(+)$. The degeneracies are split by a relatively strong magnetic field \cite{Lemke2009}, $\Delta E _{a_i} = - g_a m_i B$, where $a=e,g$ denotes the ground or excited manifold, $i=\uparrow/\downarrow$, $m_i=\pm 1/2$ the nuclear spin quantum number, $g_g/h\approx -750~$Hz/G, and $g_e/h\approx-1250$~Hz/G. A bias field $B\sim 100 ~$G thus leads to Zeeman shifts $\sim \pm 25~$kHz on the $\pi$ transitions and $\sim \pm 100~$kHz on the $\sigma^{\pm}$ transitions: these shifts are large compared to typical Rabi frequencies of both the gauge-field and probe lasers ($\sim 1$ kHz or less), and the effect of the different lasers can thus be treated independently. \\

In order to probe the $(+)$ system, one introduces a weaker additional laser, coupling $g_\uparrow \rightarrow e_\downarrow$. Due to the gauge coupling, a population will build up in the $g_\downarrow$ state as well (roughly equal to that in the $e_\downarrow$ state since $\Omega \ll J/\hbar$). Those atoms will be missing in the final detection step. After probing, the lattice sites are isolated by rapidly raising the lattice height and switching off the artificial gauge field. Atoms in the $g$ manifold are dispatched (possibly detected) using an auxiliary strong transition $g \rightarrow f$. A natural choice for $^{171}$Yb is $f \!=$$^{1}\!P_1$, with a linewidth $\gamma_f/2\pi\approx 28~$MHz much larger than any Zeeman splitting in $g$ or $e$ (thus prohibiting independent detection of atoms depending on their spin $\downarrow /  \uparrow$). Crucially, atoms in the $e$ manifold are not in resonance with the imaging light and are therefore unaffected. The $e_\downarrow$ atoms are subsequently brought down to the $g$ manifold using, {\it e.g.}, adiabatic passage techniques, leaving the $e_{\uparrow}$ state unaffected. A further imaging pulse allows to detect those atoms, initially excited by the probe pulse. One might worry that a fraction of atoms from the $e_\uparrow$ state could end up being transferred too, thus contaminating the final edge signal. Fortunately, the off-resonant excitation rate to ``wrong" states will be smaller than the resonant rate by a factor scaling as $\sim (\Omega/\Delta_Z)^2$, with $\Delta_Z\sim \vert g_e-g_g \vert B/2$ a typical Zeeman splitting. Taking for example the parameters given in \cite{gerbier2010a}, one has $J/h\approx 100~$Hz, and $\Omega \sim 0.05 J/\hbar \sim 2\pi \times 20~$Hz, making the final contamination of $g_\downarrow$ by $e_\uparrow$ negligible ($\sim 10^{-5}$).

\begin{figure}[h!]
	\centering
	\includegraphics[width=1\columnwidth]{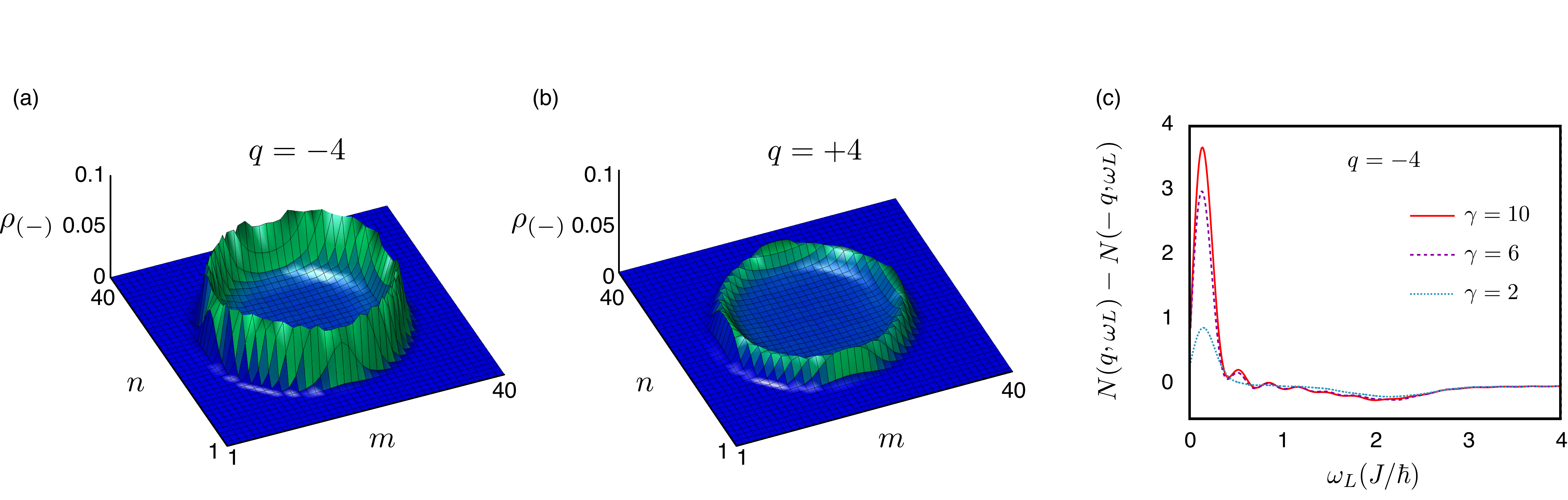}
	\caption{\label{figure9} (a)-(b) Density $\rho_{(-)} (m,n)$ for $q=\pm 4$, $\gamma=10$ and $\omega_{L} = 0.14 J/\hbar$. The probe and system parameters in (a)-(b) are the same as in Fig.\ref{figure7}. (c) Excitation fraction versus probe frequency, for several potential shapes.} 
\end{figure}

\subsection{Analysis of the Shelving method}
\label{ana}

In order to study the effect of the Shelving method, we consider a simplified level scheme with two internal states only, denoted by the indices $(\pm)$.  We suppose that only the $(+)$ sector is initially populated. The spatial profile of the coupling laser $f_L (r)$ is similar to the one used for Bragg excitations, but now the Pauli principle does not  restrict the available final states, since the state $(-)$ is initially unoccupied. We write the coupling to the probe as
\begin{align}
&\hat H_{\text{Shelving}}(t)= \hbar \Omega \bigl ( \hat{W}_q^{\text{sh}} e^{-i \omega_L t} +  \bigl (\hat{W}_{q}^{\text{sh}} \bigr)^{\dagger} e^{i \omega_L t} \bigr ) , \quad \hat{W}_q^{\text{sh}}=\sum_{\alpha \beta} I_{\alpha \beta}^q \hat c_{\alpha (-)}^{\dagger} \hat c_{\beta (+)},
\end{align}
where the operator $\hat c_{\alpha (\pm)}^{\dagger}$ creates a particle of the $(\pm)$ sector in the eigenstate $\vert \psi_{\alpha} \rangle$, and where $I_{\alpha \beta}^q$ has the same definition as in Eq. \eqref{integrale}, since $\hat H _{(-)}= \hat H _{(+)}=\hat{H}_0$.  The $(+)$ sector is initially populated, such that the initial and excited states have the following forms
\begin{align}
&\vert 0 \rangle\!=\! \vert 1 \dots 1 \underbrace{\vert}_{E_{\text{F}}} 0 \dots 0 \rangle_{(+)} \vert 0 \dots 0 \rangle_{(-)}, \label{shelve1} \\
&\vert kl \rangle\!=\! \vert 1 \dots 1 \underbrace{0}_{l} 1 \dots 1 \underbrace{\vert}_{E_{\text{F}}} 0 \dots 0 \rangle_{_{(+)}} \vert 0 \dots 0\underbrace{1}_{k} 0 \dots 0 \rangle_{_{(-)}}, \label{shelve2}
\end{align}
where we suppose that $\Omega \ll J/\hbar$ to neglect higher order excitations. We note that $k$ is no longer restricted by the Pauli principle, such that $\omega_{kl}=(\epsilon_k -\epsilon_l )/\hbar$ may now take negative values. We follow the same treatment as for the Bragg scheme and we obtain the excitation fraction as
\begin{equation}
N(q,\omega_L)  = 2 \pi \Omega^2 t  \sum_{l \le E_{\text{F}}} \sum_k \vert I_{kl}^q \vert^2 \delta^{(t)} (\omega_{kl} - \omega_L ) , \label{fermiGR2}
\end{equation}
which differs from Eq. \eqref{fermiGR} by the fact that the final states $k$ are now unrestricted.  However, we stress that the sum over the initial states, $\sum_{l \le E_{\text{F}}}$ in Eq. \eqref{fermiGR2}, is still restricted by the Pauli principle: for $\omega_L \!\ll\! J/\hbar$ and when $E_{\text{F}}\!=\!-1.5 J$, this allows to probe the edge states that are located in the first bulk gap only. This important fact leads to the asymmetry highlighted in the corresponding excited fraction $N (q, \omega_L)$, illustrated in Fig. \ref{figure9}(a)-(b), which demonstrates the specific chirality of the extracted edge states. The excitation fraction $N(q,\omega_L)-N(-q,\omega_L)$ is represented in Fig. \ref{figure9}(c), showing a clear resonance peak at low frequencies $\hbar  \omega_{L}  \ll J$. Interestingly, this result shows that the low-energy regime is still governed by the chiral edge states located in the bulk gap, although transitions are now allowed for all the states below $E_{\text{F}}$, including the bulk states. Indeed, the signal $N(q,\omega_L)-N(-q,\omega_L)$ remains small and flat in the ``edge-bulk" region, while the chiral ``edge-edge" peak stands even clearer than in the Bragg case (since more ``edge-edge" transitions are allowed between states of same chirality). By setting the probe parameters $(q, \omega_L)$ close to a resonance peak, one can now populate edge states into the $(-)$ sector and directly visualize them using state-selective imaging. The corresponding density $\rho_{(-)} (m,n)=\langle \Psi (t) \vert \hat n_{(-)} (m,n)\vert \Psi (t) \rangle$ is illustrated in Figs. \ref{figure9}(a)-(b) for $q=\pm 4$. The clear difference between the two images, obtained with different signs of $q$ but otherwise identical setups, is a direct proof of the chiral nature of the edge excitations populated by the probe. \\

All the computations presented in this work were performed by setting the Fermi energy at the specific value $E_{\text{F}}=-1.5 J$. However, we point out that the clear signature of chiral edge states, namely the ``edge-edge" resonance peak at low frequencies, does not rely on this particular value. Indeed, this clear signature remains robust as long as the Fermi energy lies within the lowest bulk gap, which can be realized by preparing an optical lattice  with central density $n \approx 1/3a^2$. Moreover, we note that a finite temperature, small compared to $\Delta$, will not affect qualitatively  our findings, since under this condition, there will always be sufficiently many occupied edge states reacting to the probe. \\

We have seen that the number of edge states $N_{\text{edge}}$ that are available within the lowest bulk gap, and which sets an upper bound for the number of extracted particles, is related to the edge state angular velocity: $N_{\rm edge}\sim J /\hbar \dot{\theta}_e $. Besides, for a fixed angular momentum transfer $q$, the location of the resonance peak $\omega^{\text{res}}_L$ scales with this same angular velocity: $\omega^{\text{res}}_L \sim q \dot{\theta}_e$. Therefore, the Bragg spectrum resulting from a system that features a large number of edge states, such as a harmonically trapped gas $\gamma =2$, will necessarily show an ``edge-edge" resonance peak at very small frequencies, since $$  \omega^{\text{res}}_L \sim J q / \hbar N_{\text{edge}}. $$ However, due to the broadening of the resonance peak for finite times, it is desirable that this peak be centered around reasonably high frequencies,  in order to clearly distinguish between $N(q, \omega_L) \ne N(-q, \omega_L)$. Therefore, one has to make a compromise between detecting a large number of atoms per probe pulse and limiting the effects of finite time broadening, by designing a system with a reasonably large number of available edge states with sufficiently large angular velocity. Predicting the optimal value of $\gamma$ would require a precise knowledge of experimental numbers, which could vary from one experiment to the other. \\


We finally stress that the condition considered for the shelving method $$\hat H _{(-)}\!=\! \hat H _{(+)}\!=\!\hat{H}_0,$$ is necessary in order to probe the edge-state structure. Indeed, if we consider a simpler scheme in which the $(-)$ sector is no longer subjected to a synthetic gauge potential, we find that $N(q,\omega_L)-N(-q,\omega_L) \approx 0$. This observation shows that our scheme requires that the edge states of the $(-)$ sector should have the same chirality than the initially populated edge-states of the $(+)$ sector, \emph{i.e.} both systems should be subjected to the same synthetic magnetic flux. \\

\section{Conclusion}

In this work, we showed that cold atoms trapped in optical lattices and subjected to synthetic magnetic fields offer a unique platform to investigate the physics of topological edge states. In particular, we showed that an elegant detection scheme, referred to as the shelving  method, allows to identify and extract topological edge states in a highly controllable way, but also to image them using available imaging technics. \\
In this sense, a cold-atom simulation of the quantum Hall setup, together with our detection scheme, offers the possibility to directly visualize the edge states and to access their dispersion relation, without performing any transport measurements. In particular, such an experiment could be realized in the interacting regime, where the dispersion relation of fractional QH edge states could be accessed. \\
Let us stress that our detection method does not rely on the lattice, nor on the setup which generates the synthetic
magnetic field (e.g. laser-induced \cite{aidelsburger2011a}, lattice shaking \cite{struck2012a}, atom-chip \cite{Goldman2010a}, rotation, ...). Consequently, our method applies for a wide range of cold-atom realizations of 2D topological phases, displaying a circular geometry. In particular, exploiting atomic species with many internal states, one could easily extend the present detection scheme to probe the physics of helical edge states  exhibiting the quantum spin Hall effect \cite{HasanKane2010,Goldman2010a,Goldman:2012epl,Hauke:2012}.  We finally mention the possibility of using ``photon-counting"  techniques as an alternative method for detecting the Bragg excitations resulting from our Laguerre-Gauss  probing lasers \cite{Pino:2011}.

\paragraph{Acknowledgments}
We acknowledge the F.R.S-F.N.R.S, DARPA (Optical lattice emulator project), the Emergences program (Ville de Paris and UPMC)  and ERC (Manybo Starting Grant) for financial support. N.G. thanks the organizers of the Lyon BEC 2012 Conference and Ian B. Spielman for discussions.


\end{document}